\documentclass[final,5p,times, twocolumn]{elsarticle}

\usepackage{amssymb}
\usepackage{amsmath,bm}
\usepackage{subfig,tabularx, multirow,lmodern,makecell, booktabs, amssymb,makecell}
\usepackage[ruled,vlined]{algorithm2e}
\usepackage{mathtools}
\usepackage{rotating}
\journal{ }
\usepackage{hyperref}
\usepackage{makecell}
\begin{document}

\begin{frontmatter}

\title{PAT: Privacy-Preserving Adversarial Transfer for Accurate, Robust and Privacy-Preserving EEG Decoding}

\author[1,2,3]{Xiaoqing~Chen}{}
\author[1,2]{Tianwang~Jia}{}
\author[1,2]{Yunlu~Tu}{}
\author[1,2,3]{Dongrui~Wu\corref{cor1}}
\cortext[cor1]{Corresponding author: drwu09@gmail.com}

\affiliation[1]{organization={Hubei Key Laboratory of Brain-inspired Intelligent Systems, School of Artificial Intelligence and Automation, Huazhong University of Science and Technology}, city={Wuhan}, postcode={430074}, country={China}}
\affiliation[2]{organization={Shenzhen Huazhong University of Science and Technology Research Institute}, city={Shenzhen}, postcode={518000}, country={China}}
\affiliation[3]{organization={Zhongguancun Academy}, city={Beijing}, postcode={100094}, country={China}}

\begin{abstract}
An electroencephalogram (EEG)-based brain-computer interface (BCI) enables direct communication between the brain and external devices. However, such systems face at least three major challenges in real-world applications: limited decoding accuracy, poor robustness, and privacy risks. Although prior studies have addressed one or two of these issues, methods that simultaneously improve accuracy, robustness, and privacy remain largely unexplored. In this paper, we propose Privacy-preserving Adversarial Transfer (PAT), a unified training framework that combines data alignment, adversarial training, and privacy-preserving transfer. PAT provides a single pipeline that can be instantiated under three privacy-preserving scenarios, i.e., centralized source-free transfer, federated source-free transfer, and transfer with privacy-preserved source data, while jointly improving accuracy and robustness. Experiments on five public EEG datasets under three privacy-preserving scenarios (centralized source-free transfer, federated source-free transfer, and transfer with privacy-preserved source data) show that PAT outperforms over ten classic and state-of-the-art methods in both accuracy and robustness. PAT also outperformed leading transfer learning approaches that do not incorporate any privacy mechanisms by 9.76\% in terms of average accuracy and robustness. To our knowledge, this is the first approach that simultaneously addresses all three major challenges in EEG-based BCIs. We believe this work can help motivate further research on more accurate, robust, and privacy-preserving EEG decoding.
\end{abstract}


\end{frontmatter}

\section{Introduction}\label{sec1}
A brain-computer interface (BCI) creates a direct link between the human brain and a computer, enabling the assistance, enhancement, or restoration of cognitive sensory-motor functions~\cite{Ienca2018}. BCIs have broad applications in neurological rehabilitation~\cite{Daly2008}, touch-based exploration~\cite{ODoherty2011}, robot control~\cite{Hochberg2012}, speech synthesis~\cite{Metzger2023}. Depending on electrode placement, BCIs can be grouped into non-invasive, partially invasive, and invasive systems. Non-invasive BCIs, which often use scalp electroencephalogram (EEG) signals~\cite{NicolasAlonso2012} as input, are the most widely used due to their ease of deployment.

Despite advantages such as non-invasiveness and low cost, real-world EEG-based BCIs still face several critical challenges that limit their practical deployment.

First, EEG decoding models often suffer from \emph{limited accuracy} in real-world scenarios~\cite{drwuMITLBCI2022}. The performance of machine learning models relies heavily on high-quality and sufficiently large datasets. However, collecting EEG signals is time-consuming and labor-intensive, and EEG data are highly non-stationary: recordings across equipment, sessions, or users can vary significantly. Such limited data availability and substantial inter-domain variability greatly hinder further improvements in EEG decoding accuracy.

Second, EEG decoding systems typically exhibit \emph{poor robustness} against various perturbations. Robustness is essential for reliable BCI applications, especially in safety-critical scenarios. Unfortunately, recent studies have shown that BCI systems are vulnerable to both adversarial attacks and random noise~\cite{Zhang2019, Zhang2020, Wang2022, Jung2023}. Imperceptible adversarial perturbations may arbitrarily manipulate the model output, and even random noise may also degrade the performance dramatically. These vulnerabilities severely restrict the practical use of BCIs in clinical and daily-life applications.

Third, EEG-based BCIs involve serious \emph{privacy risks} that are often overlooked. User privacy is strictly regulated by laws such as the European Union General Data Protection Regulation and the China Personal Information Protection Law. However, EEG signals inherently contain sensitive personal information, including personality traits, cognitive abilities, and private emotional states, regardless of the experimental paradigm~\cite{Landau2020,Zhang2021a,Xia2023}. Ensuring effective privacy protection is therefore crucial for the trustworthy and widespread adoption of BCI systems.

\begin{figure}[htpb]\centering
	{\includegraphics[width=1\linewidth,clip]{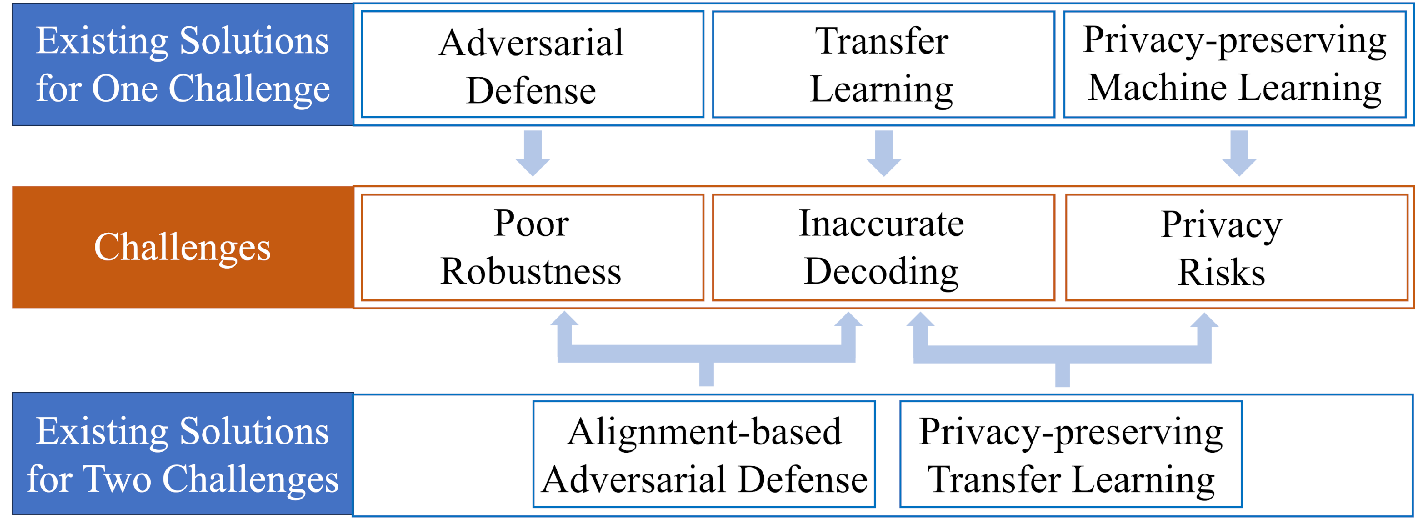}}
	\caption{Five categories of existing studies for addressing the three challenges in BCIs.} \label{fig:existing}
\end{figure}

Extensive research has been conducted to address these challenges and facilitate practical BCI applications. According to the specific problems they target, existing studies can be grouped into five categories: transfer learning, adversarial defense, privacy-preserving machine learning, alignment-based adversarial defense, and privacy-preserving transfer learning. These categories are summarized in Figure~\ref{fig:existing}.

Transfer learning leverages data from different domains to improve EEG decoding in a target domain~\cite{He2019, drwuMITLBCI2022}. Adversarial detection and robust training, i.e., adversarial defense methods, enhance the robustness of BCIs against perturbations~\cite{Meng2023a, Chen2024b}. Privacy-preserving machine learning and data perturbation techniques help protect the privacy of EEG data~\cite{Liang2022, Chen2024a}. However, these approaches typically address only one of the three challenges. Tackling multiple challenges simultaneously is more difficult, and prior work has shown that improving one aspect may degrade performance in others. For example, cross-subject transfer learning may expose the privacy of source users, while adversarial defense may reduce classification accuracy on benign samples~\cite{Meng2023a}.

Some recent studies have attempted to address two of these challenges simultaneously. For example, alignment-based adversarial training (ABAT)~\cite{Chen2024,Chen2025} combines data alignment and robust training to jointly improve the accuracy and adversarial robustness of EEG decoding models. Privacy-preserving transfer learning~\cite{Xia2022, Liang2022} improves classification accuracy while protecting the privacy of source data. However, ABAT does not exploit EEG data or decoding models from other domains to further enhance target-domain decoding performance. Privacy-preserving transfer learning, in turn, does not consider model robustness. It remains unclear whether enhancing robustness within privacy-preserving transfer learning would inevitably degrade decoding accuracy.

In this work, we propose Privacy-preserving Adversarial Transfer (PAT), a general framework that unifies data alignment, adversarial training, and privacy-preserving transfer into a single training pipeline. Instead of designing separate algorithms for each setting, PAT provides a common procedure that can be instantiated in three representative privacy-preserving scenarios, i.e., centralized source-free transfer, federated source-free transfer, and transfer with privacy-preserved source data, while jointly enhancing accuracy and robustness.

Experiments on five public EEG datasets under these three scenarios validate the effectiveness of PAT and demonstrate the feasibility of accurate, robust, and privacy-preserving EEG decoding. In all three privacy-preserving scenarios, PAT simultaneously improves the accuracy and robustness of the target user's EEG decoding model.

In summary, our main contributions are:
\begin{enumerate}
	\item We propose PAT, a unified training framework that jointly integrates data alignment, adversarial training, and privacy-preserving transfer for EEG-based BCIs. Within this framework, we demonstrate that accurate, robust, and privacy-preserving EEG decoding can be achieved simultaneously, facilitating practical BCI deployment.
	\item We instantiate PAT in three representative privacy-preserving scenarios, i.e., centralized source-free transfer, federated source-free transfer, and transfer with privacy-preserved source data, showing that the same framework can be adapted to different privacy constraints by only changing the accessible source information (models or perturbed data).
	\item We compare PAT with over ten state-of-the-art transfer learning approaches on five EEG datasets across three privacy-preserving transfer learning scenarios, demonstrating its superior performance. PAT even surpass leading transfer learning approaches that do not incorporate any privacy mechanisms.
\end{enumerate}

\section{Related work}\label{app:rw}

This section reviews related work on transfer learning, adversarial defense, and privacy protection.

\subsection{Transfer learning}

Transfer learning leverages data from source domains to improve learning in a target domain. A central challenge is to reduce distribution discrepancies between source and target domains, typically in terms of marginal and/or conditional probability distributions.

Marginal distribution discrepancy can be reduced by input transformations (e.g., Euclidean Alignment (EA)~\cite{He2019}), distribution alignment (e.g., Deep Adaptation Network (DAN)~\cite{Long2019}, Multi-dataset joint pre-Traing (mdJPT)\cite{Zhang2025}, Alignment-based Frame Patch Modeling (AFPM)~\cite{Chen2025a}), or adversarial training (e.g., Domain-Adversarial Neural Network (DANN)~\cite{Ganin2016}). Conditional discrepancy is often reduced via pseudo-labeling (e.g., Joint Adaptation Networks (JAN)~\cite{Long2017}) or uncertainty minimization (e.g., Minimum Class Confusion (MCC)~\cite{Jin2020}, Source Hypothesis Transfer (SHOT)~\cite{Liang2022}). Margin Disparity Discrepancy (MDD)~\cite{Zhang2019b} and Conditional Domain Adversarial Networks (CDAN)~\cite{Long2018} jointly reduce both marginal and conditional discrepancies.

Transfer learning approaches can substantially improve the decoding accuracy for a target user in EEG applications, yet they generally overlook the protection of source users' private information.

\subsection{Adversarial defense}

A variety of adversarial defense methods have been developed in computer vision and natural language processing~\cite{Madry2018}. Among them, robust training~\cite{Chen2022}, which augments the training set with adversarial examples, is one of the most effective strategies. Adversarial training~\cite{Madry2018} is a widely used robust training approach, with many variants~\cite{Zhang2019a,Zhang2021}.

Adversarial defense for BCIs has also attracted growing attention~\cite{Xue2022, Wu2023, Meng2023a, Chen2024}. Li \emph{et al.}~\cite{Li2022} evaluated five adversarial-training-based defense methods for BCIs. Meng \emph{et al.}~\cite{Meng2023a} provided a comprehensive analysis of the advantages and limitations of different adversarial defense strategies in BCIs. Gunawardena \emph{et al.}~\cite{Gunawardena2024} combined adversarial training with adversarial detection to defend against black-box attacks in EEG-based BCIs. However, prior studies have shown that although adversarial defenses improve robustness, they often reduce classification performance on benign samples~\cite{Li2022, Meng2023a}. To alleviate this trade-off, Chen \emph{et al.}~\cite{Chen2024, Chen2025} proposed ABAT for EEG-based BCIs, which performs adversarial training on aligned EEG data and improves accuracy on both benign and adversarial samples.

Compared with existing methods, PAT is designed as a framework rather than a single algorithm. ABAT focuses on combining alignment and adversarial training within a single domain, without leveraging source priors across. In contrast, PAT provides a unified procedure that can incorporate either source models or privacy-preserved source data, and adapts seamlessly across centralized, federated, and perturbed-data scenarios. PAT extends ABAT in many aspects, as shown in Table~\ref{tab:com}.

\renewcommand{\arraystretch}{1.3}
\begin{table*}[htbp]
\small   \centering
  \caption{Comparisons between ABAT and PAT.}
    \begin{tabular}{lll}
    \toprule
    Feature & ABAT & PAT (This paper) \\
    \midrule
    Research Scope & Single-domain EEG decoding & Cross-domain EEG decoding\\
    Addressed Challenges & Accuracy and robustness & Accuracy, robustness, and privacy protection  \\
    Source Prior Integration & None & Integrate source models or privacy-preserved data \\
    \makecell[l]{Privacy-Preserving Mechanisms} & None & \makecell[l]{Federate or centralized source-free transfer,\\ user-specific perturbation} \\
    \multirow{2}[0]{*}{Training Pipeline} & \multirow{2}[0]{*}{EA $\rightarrow$ AT } & EA $\rightarrow$ Data Augmentation $\rightarrow$  \\
      &   & AT + Source Prior Guidance \\
    Applicable Scenarios & Single-domain scenarios & Three privacy-preserving cross-domain scenarios \\
    \bottomrule
    \end{tabular}%
  \label{tab:com}%
\end{table*}%
\renewcommand{\arraystretch}{1}

\subsection{Privacy protection}

Various privacy-preserving methods have been proposed for EEG-based BCIs to prevent the leakage of sensitive information from EEG data. These methods can be broadly divided into privacy-preserving machine learning based approaches and data perturbation based ones~\cite{Xia2023}.

Privacy-preserving machine learning avoids directly sharing raw EEG data or model parameters. Typical approaches include source-free transfer learning~\cite{Xia2022, Zhang2022, Zhang2023} and federated learning~\cite{Jia2024, Wei2025, Jiang2025, Liu2026, Gupta2025, MartinezBeltran2026}, where training data remain local, thus preserving data privacy. For example, Zhang \emph{et al.}~\cite{Zhang2022} investigated gray-box and black-box transfer for MI and affective BCIs, and Jia \emph{et al.}~\cite{Jia2024} restricted the transfer of client data to a central server via federated learning.

Data perturbation, which adds perturbations to the original data to prevent models from learning private information while preserving utility for downstream tasks. For instance, Meng \emph{et al.}~\cite{Meng2023} designed sample-specific and user-specific perturbations to generate identity-unlearnable EEG data, and Chen \emph{et al.}~\cite{Chen2024a} further proposed four types of robust user-wise perturbations to protect privacy in EEG data.

Most privacy-preserving transfer methods optimize model accuracy under a fixed privacy mechanism, whereas PAT is designed to accommodate multiple privacy mechanisms within a unified training template.

\section{Materials and methods}

This section describes problem formulation and the methodology of proposed PAT.

\subsection{Problem statement}

Consider a source EEG training dataset $\mathcal{D}_S = \left\{ \left( X_{S,i}, y_{S,i}, u_i \right) \right\}_{i=1}^{N_S}$ consisting of $N_S$ samples, where $X_{S,i} \in \mathbb{R}^{c \times t}$ is the $i$-th EEG trial with $c$ channels and $t$ time-domain samples, $y_{S,i} \in \{1, \ldots, K\}$ is the corresponding BCI task label (e.g., left/right hand MI), and $u_i \in \{1, \ldots, U\}$ denotes the user identity label. Additionally, we have a calibration dataset $\mathcal{D}_T = \left\{ \left( X_{T,i}, y_{T,i} \right) \right\}_{i=1}^{N_T}$ with $N_T$ samples, where each $X_{T,i} \in \mathbb{R}^{c \times t}$ is an EEG trial with $c$ channels and $t$ time-domain samples, and $y_{T,i} \in \{1, \ldots, K\}$ is the corresponding BCI task label.

$\mathcal{D}_S$ contains both task-relevant and private information of the source subjects~\cite{Chen2024a}, so its direct access is restricted to safeguard user privacy. Instead, we only have access to a task model $\emph{C}_{\bm{\theta}}$ trained on $\mathcal{D}_S$, or a perturbed dataset $\tilde{\mathcal{D}_S}$ that masks subject-specific information of $\mathcal{D}_S$ in training the target domain classifier.

Depending on how the source information is provided under privacy constraints, we consider three representative scenarios: centralized source-free transfer, federated source-free transfer, and transfer with privacy-preserved source data.

%
\subsection{PAT}

This subsection describes the detailed pipeline of proposed PAT.

\subsubsection{Overall framework}

We instantiate PAT in three privacy-preserving scenarios. In all three privacy-preserving scenarios, PAT follows the same training pipeline. Given a small labeled calibration set from the target user and some form of source prior (either a pre-trained source model or privacy-preserved source data), PAT (1) aligns the target data distribution via Euclidean Alignment, (2) augments the target data to increase diversity, and (3) performs adversarial training on the aligned and augmented target data, optionally combined with supervised learning on source data when available.

 This leads to a unified objective that combines a robust loss on target adversarial examples with a task loss on source priors. The specific scenario only determines which source prior is available (a centralized model, a federated model, or perturbed source data), while the overall training framework remains unchanged.

\begin{figure*}[htpb]\centering
	{\includegraphics[width=0.9\linewidth,clip]{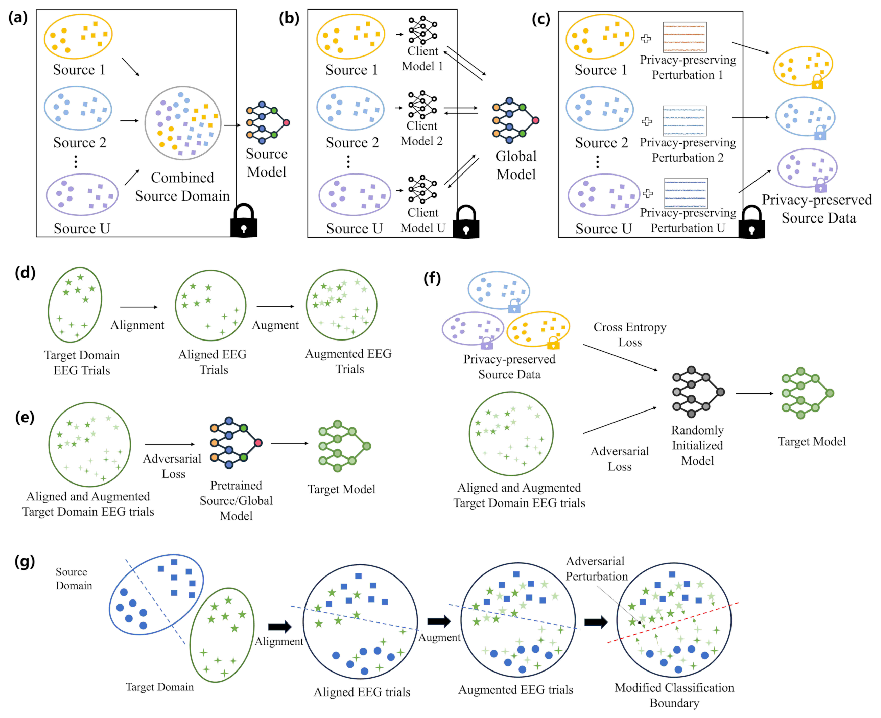}}
	\caption{Overview of PAT framework. (a) Centralized source-free model training. (b) Federated source-free model training. (c) Source data perturbation for privacy protection. (d) Target domain alignment and augment. (e) Source model calibration on target domain. (f) Joint training on privacy-preserved source data and the aligned and augmented target domain data. (g) Model classification boundary optimization process in PAT. } \label{fig:frame_our}
\end{figure*}

In centralized source-free transfer learning scenario, as shown in Figure~\ref{fig:frame_our}a, only models trained on source data can be shared. In federated source-free transfer learning scenario shown in Figure~\ref{fig:frame_our}b, source data from different domains cannot be combined together for model training. Transfer learning with privacy-preserved source domain data is shown in Figure~\ref{fig:frame_our}c, where subject-specific information is masked with perturbations.

PAT aligns and augments target domain calibration data, as illustrated in Figure~\ref{fig:frame_our}d. Either task model $\emph{C}_{\bm{\theta}}$ trained in the source domain or perturbed privacy-preserved source data $\tilde{\mathcal{D}}_S$ can be used to facilitate the training of EEG decoding models in the target domain in PAT, as shown in Algorithm~\ref{Alg:ppn}. When solely utilizing source domain models, PAT implements adversarial training on the aligned and augmented target domain data to optimize the model, as demonstrated in Figure~\ref{fig:frame_our}e. In scenarios where privacy-preserved source data is available, PAT jointly trains models using both the privacy-preserved source data and the aligned and augmented target domain data, as shown in Figure~\ref{fig:frame_our}f.

The source domain models/data inherently provide a good prior to guide target domain EEG decoding. PAT further enhances decoding accuracy and robustness through adversarial training on aligned and augmented target domain data, as depicted in Figure~\ref{fig:frame_our}g.

\begin{algorithm}[htpb]
	\caption{Algorithm for PAT.}\label{Alg:ppn}
	\KwIn{$\emph{C}_{\bm{\theta}}$, task classifier trained in the source domain\;
		\hspace*{10mm} $\tilde{\mathcal{D}}_S$, perturbed privacy-preserving source data\;
		\hspace*{10mm} $\mathcal{D}_T$, target domain calibration EEG data\;}
	
	\KwOut{$\emph{C}^{\prime}_{\bm{\theta}}$, target classifier.}

	Perform data alignment on $\mathcal{D}_T$ by (\ref{alg:ea1}) and (\ref{alg:ea2})\;
	
	Perform data augmentation on $\mathcal{D}_T$ by (\ref{alg:aug}) to obtain augmented calibration set $\mathcal{D}'_T$\;
	
	\tcp{Centralized source-free transfer learning scenario, or federated source-free transfer learning scenario}
	Fine-tune $\emph{C}_{\bm{\theta}}$ on $\mathcal{D}'_T$ by (\ref{algo:at}), (\ref{pgd1}) and (\ref{pgd2}) to obtain target classifier $\emph{C}^{\prime}_{\bm{\theta}}$\;
	
	\tcp{Transfer learning scenario with privacy-preserved source domain data}
	
	Train target classifier $\emph{C}^{\prime}_{\bm{\theta}}$ on $\mathcal{D}'_T$ and $\tilde{\mathcal{D}}_S$ by (\ref{pgd1}), (\ref{pgd2}) and (\ref{algo:atc}).
\end{algorithm}

\subsubsection{Centralized source-free transfer}

Centralized source-free transfer is a special case of PAT where the source prior is a single model trained on centrally aggregated source data. In this scenario, the source subjects trust each other, so they can share raw EEG data, but not with the target user. EEG data from multiple source users are aggregated to train a single source model, which is then utilized in transfer learning to train the target model, as illustrated in Figure~\ref{fig:frame_our}a. The source data privacy is protected because the target user cannot access the source domain data \cite{Xia2022}.

\subsubsection{Federated source-free transfer}

Federated source-free transfer instantiates PAT with a global model obtained via federated learning, which serves as the source prior. In this more strict scenario, no source user wants to share his/her raw EEG data with other source users and the target user. Without access to the source users' private EEG data, a central server maintains a global model and distributes it to each source user for updates. Each source user refines the global model parameters using his/her own EEG data and sends these updates back to the server for aggregation, as illustrated in Figure~\ref{fig:frame_our}b. FedBS~\cite{Jia2024}, which achieves superior performance compared to centralized training, is utilized in this paper.

Assume there are $S$ clients, and the $s$-th client has $n_s$ EEG trials and a local classifier $C_s$. The global model $C$ is obtained by:
\begin{align}
	C=\sum_{s=1}^{S} \frac{n_s}{\sum_{s=1}^{S} n_s} C_s.
\end{align}
This global model $C$ is then used for target model training.

\subsubsection{Transfer with privacy-protected source data}

A user-specific perturbation can be applied to each source domain to safeguard source user privacy while leveraging the source data for target domain classifier training, as illustrated in Figure~\ref{fig:frame_our}c. This perturbation is designed to make user identity information effectively unlearnable (i.e., prevent reliable identity classification) while preserving BCI task-relevant information.

Given a source EEG training dataset $\mathcal{D}_S = \left\{ \left( X_{S,i}, y_{S,i}, u_i \right) \right\}_{i=1}^{N_S}$, we generate a privacy-unlearnable EEG training dataset $\tilde{\mathcal{D}}_S=\left\{\left(\tilde{X}_{S,i}, y_{S,i}, u_i\right) \right\}_{i=1}^{N_S}$, by adding a user-wise perturbation $\Delta_{u_i}\in  \mathbb{R}^{c \times t}$ to each EEG trial $X_{S,i}$, i.e., $\tilde{X}_{S,i}=X_{S,i} + \Delta_{u_i}$. We used the synthetic noise approach proposed in~\cite{Chen2024a} to calculate $\Delta_{u_i}$.

\subsubsection{Calibration data alignment and augmentation}

Given the target domain calibration set $\mathcal{D}_T = \left\{ \left( X_{T,i}, y_{T,i} \right) \right\}_{i=1}^{N_T}$. PAT has the following two steps:
\begin{enumerate}
	\item \emph{Data alignment}, where EA is performed on the calibration dataset $\mathcal{D}_T$. EA first computes the Euclidean mean $\bar{R}$ of all $N_T$ spatial covariance matrices:
	\begin{align}
		\bar{R}=\frac{1}{N_T} \sum_{i=1}^{N_T} X_{T,i}\left(X_{T,i}\right)^{\top}.\label{alg:ea1}
	\end{align}
	It then aligns each trial by:
	\begin{align}
		X_{T,i}\leftarrow\bar{R}^{-1 / 2} X_{T,i}, \quad i=1,...,N_T. \label{alg:ea2}
	\end{align}
	
	After EA, the EEG trials are whitened, i.e., the average spatial covariance matrix becomes the identity matrix. Since EA is also performed in each source domain, EEG data distributions from different users become more consistent. Note that other alignment approaches may also be used~\cite{Pan2025}.
	
	\item \emph{Data augmentation}, which increases the diversity of the calibration data. We use a simple scaling-based data augmentation that multiplies the amplitude of an EEG trial $X$ by a coefficient close to 1~\cite{Freer2020}:
	\begin{align}
		X'=X\cdot(1\pm \beta), \quad X\in \mathcal{D}_T.\label{alg:aug}
	\end{align}
	$\beta=0.05$ was used in our experiments, according to \cite{Freer2020, Chen2025}. After data augmentation, we obtain the augmented calibration set $\mathcal{D}'_T=\left\{\left(X_{T,i}, y_{T,i}\right),\left(X'_{T,i}, y_{T,i}\right)\right\}_{i=1}^{N_T}$. Note that other data augmentation approaches may also be used~\cite{Pan2025}.
\end{enumerate}

\subsubsection{Adversarial transfer with source model}

Adversarial training generates and uses adversarial samples in training to enhance the classifier robustness and generalization. It can be formulated as a min-max (saddle point) optimization problem:
	\begin{align}
		\min_{\bm{\theta}} \mathbb{E}_{(X_{T}, y_{T}) \sim \mathcal{D}'_T}\left[\max_{X^{adv}_T \in \mathcal{B}(X_{T}, \epsilon)} \mathcal{L}\left(C_{\bm{\theta}}\left(X^{adv}_{T}\right), y_{T}\right)\right],\label{algo:at}
	\end{align}
	where $\mathcal{B}(X_{T}, \epsilon)$ is the $\ell_\infty$ ball of radius $\epsilon$ centered at $X_{T}$, $\emph{C}_{\bm{\theta}}$ a classifier with parameter $\bm{\theta}$, and $\mathcal{L}$ its loss function. In this paper, $X^{adv}$ was an adversarial sample generated by projected gradient descent (PGD)~\cite{Madry2018}. PGD starts from a perturbed version $X^{adv,0}$ of a benign sample $X$:
	\begin{align}
		X^{adv,0} = X + \bm{\xi},\label{pgd1}
	\end{align}
	where $\bm{\xi}$ is uniform random noise sampled from $(-\epsilon,\epsilon)$. $X^{adv,i}$ is then iteratively updated by:
	\begin{equation}
		\begin{aligned}[b]
			X^{adv,i} = & \text{Proj}_{X, \epsilon} \Biggl[
			X^{adv,i-1} \\  &+ \alpha \cdot \text{sign} \biggl( \nabla_{X^{adv,i-1}} \mathcal{L} \Bigl( C_{\bm{\theta}}(X^{adv,i-1}), y \Bigr) \biggr) \Biggr],
		\end{aligned}
		\label{pgd2}
	\end{equation}
	where $\alpha \leq \epsilon$ is the step size. The function $\text{Proj}_{X, \epsilon}$ ensures that $X^{adv,i}$ remains within the $\epsilon$-neighborhood of $X$ according to the $\ell_\infty$ norm.
	
\subsubsection{Adversarial transfer with privacy-protected source data}
	When a privacy-unlearnable EEG training dataset $\tilde{\mathcal{D}}_S=\left\{\left(\tilde{X}_{S,i}, y_{S,i}, u_i\right) \right\}_{i=1}^{N_S}$ is given, the optimization problem becomes:
	\begin{equation}
		\begin{split}
			\min_{\bm{\theta}} \mathbb{E}_{\substack{(X_{T}, y_{T}) \sim \mathcal{D}'_T,\\(\tilde{X}_{S}, y_{S}) \sim \tilde{\mathcal{D}}_S}}
			\Biggl[ \max_{X^{adv}_T \in \mathcal{B}(X_{T}, \epsilon)}
			&\mathcal{L}\left(C_{\bm{\theta}}\left(X^{{adv}}_{T}\right), y_{T}\right) \\
			&+ \mathcal{L}\left(C_{\bm{\theta}}\left(\tilde{X}_{S}\right), y_{S}\right) \Biggr],
		\end{split}
		\label{algo:atc}
	\end{equation}
	where the model is optimized on the benign samples of the source domain and the adversarial samples of the target domain.

\section{Experiments and results}\label{sect:results}

This section presents experimental results to demonstrate the effectiveness of our proposed approach.

\subsection{Datasets}

Our experiments were conducted on five EEG datasets, including three motor imagery (MI) datasets, one seizure detection dataset, and one emotion recognition dataset. A summary of these datasets is provided in Table~\ref{tab:datasets}:
\begin{enumerate}
\item BNCI2014001~\cite{Tangermann2012}, i.e., Dataset 2a of BCI Competition IV, contains data from 9 subjects performing four-class MI (left hand, right hand, both feet, and tongue). The 22-channel EEG signals were sampled at 256 Hz, and each subject completed 144 trials per class.
\item Weibo2014~\cite{Yi2014} includes data from 10 subjects performing seven-class MI (left hand, right hand, both feet, both hands, left hand combined with right foot, right hand combined with left foot, and rest). Data collection was divided into 9 sessions with 5-10 minute breaks between sessions. In this study, only the first six classes were used. The 64-channel EEG signals were sampled at 200 Hz, and each subject completed 80 trials per class.
\item BNCI2014002~\cite{Steyrl2016} contains data from 14 subjects performing left- and right-hand MI. The 15-channel EEG signals were sampled at 250 Hz, and each subject completed 80 trials per class.
\item NICU~\cite{Stevenson2019} is a seizure detection dataset with 18 bipolar EEG channels from 79 subjects. Each recording is approximately 74 minutes long and was annotated every second as epileptic or normal by three experts. In this paper, 8 subjects (S13, S21, S31, S34, S36, S62, S66, and S75), whose annotations were highly consistent across the three experts, were selected. Ground-truth seizure labels were obtained via majority voting among the three experts.
\item SEED~\cite{Zheng2015} is an EEG-based emotion recognition dataset collected from 15 subjects using a 62-channel EEG cap. Three emotions (negative, positive, and neutral) were elicited using 15 film clips. Eighteen emotion-related channels were used in our experiments: FP1, FPZ, FP2, FZ, CZ, PZ, F3, F4, AF3, AF4, C3, C4, P3, P4, O1, O2, T7, and T8.
\end{enumerate}

For the three MI datasets, EEG data from [0, 4] seconds after each imagination cue were extracted, band-pass filtered in [8, 32] Hz, and resampled to 128 Hz. For NICU, a 50 Hz notch filter and a [0.5, 50] Hz band-pass filter were applied; for SEED, a [1, 50] Hz band-pass filter was used. The EEG recordings of NICU and SEED were then segmented into non-overlapping 4-second trials and resampled to 128 Hz.

\begin{table}[htbp] \centering \setlength{\tabcolsep}{1mm}\scriptsize
	\caption{Summary of the five datasets.}\label{tab:datasets}
	\begin{tabular}{c|ccccc} \toprule
		\multirow{2}{*}{Dataset}&\# &\# Time&\# & \# Trials &\#\\
		&Subjects&Points&Channels&per Subject&Classes\\ \midrule
		BNCI2014001&9&512&22&576&4\\
		Weibo2014&10&512&64&480&6\\
		BNCI2014002&14&512&15&160&2\\
		NICU&8&512&18&1200-5400&2\\
		SEED&15&512&18&842&3\\ \bottomrule
	\end{tabular}
\end{table}

\subsection{Evaluation metrics}

We evaluate performance using classification accuracies on benign, adversarial, and noisy samples. PGD~\cite{Madry2018}, calculated by (\ref{pgd1}) and (\ref{pgd2}), was used to generate adversarial samples.

Given an EEG trial $X$, the noisy sample $X'$ was calculated as:
\begin{align}
	X' = X + \eta \cdot \boldsymbol{\sigma}(X)\cdot \mathrm{U}(-1,1),\label{algo:noise}
\end{align}
where $\mathrm{U}(-1,1)\in\mathbb{R}^{c \times t}$ is a noise matrix whose each element is a uniformly distributed number in $[-1,1]$. $\boldsymbol{\sigma}(X)$ denotes the vector composed of the standard deviations of each channel in $X$:
\begin{align}
	\boldsymbol{\sigma}(X) = \left[\sigma(X(1)), \sigma(X(2)), \ldots, \sigma(X(c))\right]',
\end{align}
where $\sigma(X(1))$ denotes the standard deviation of the first channel in $X$.

To comprehensively evaluate the robustness of the classifiers, we computed their mean accuracies on adversarial and noisy samples with different perturbation magnitudes. For adversarial/noisy samples, we computed the classification accuracies on samples with adversarial perturbation magnitudes $\epsilon=\{0.01, 0.03, 0.05\}$/$\eta=\{1, 2, 3\}$ times the standard deviation of the original sample channels. We repeated each experiment five times, and report their averages.

\begin{figure*}[htbp]\centering

	{\includegraphics[width=0.99\linewidth,clip]{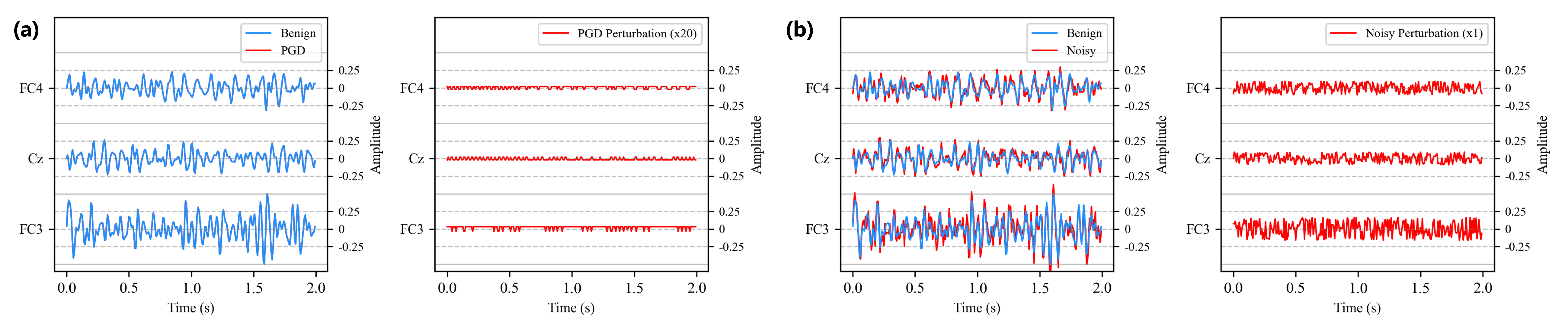}}
	\caption{A benign EEG sample and the corresponding (a) PGD adversarial sample; and, (b) Noisy sample. Note that the PGD perturbation are so small that the benign and PGD samples almost overlap. Although the adversarial sample closely resembles the benign one, it can degrade EEG classification performance greatly.} \label{fig:perturb}
\end{figure*}

Figure~\ref{fig:perturb} shows a benign EEG sample, its PGD counterpart with perturbation magnitude $\epsilon=0.01$, and its noisy counterpart with perturbation magnitude $\eta=1$. The adversarial perturbation and the random noise differ significantly. Although the adversarial sample closely resembles the benign sample, it can lead to a substantial drop in EEG classification performance, as shown in Tables~\ref{tab:main} and~\ref{tab:appnicuseed}.

\subsection{Baseline algorithms}

We compared our proposed PAT with the following algorithms from the literature:
\begin{enumerate}
\item Classical deep/machine learning approaches, i.e., common spatial pattern (CSP)~\cite{Blankertz2008} and EEGNet~\cite{Lawhern2018}. CSP is a standard MI classification method that uses labeled data to design spatial filters, perform feature extraction, and apply Linear Discriminant Analysis (LDA) for classification. The labeled data can come from the source domain only (S), the target domain only (T), or the combination of source and target domains (S\&T). EEGNet is a compact end-to-end convolutional neural network for EEG decoding. The labeled source and target data are combined during mini-batch gradient descent training. One variant samples each mini-batch randomly (typically with more source than target samples), referred to as S\&T. Another variant samples half of each mini-batch from each domain and then combines them into a full mini-batch, ensuring an equal number of samples from both domains; this variant is referred to as S+T.

\item Alignment-based robust training approaches, which leverage data alignment and adversarial training to simultaneously improve model accuracy on both benign and adversarial samples~\cite{Chen2024}. We adopted adversarial defense methods with diverse defense strategies, including AT~\cite{Madry2018}, TRADES~\cite{Zhang2019a}, SCR~\cite{Zhang2021}, AWP~\cite{Wu2020}, SEAT~\cite{Wang2022a}, RSE~\cite{Liu2018}, IT~\cite{Meng2023a}, and SAP~\cite{Dhillon2018}.

\item Source-free transfer learning approaches, which fine-tune source models trained on source data. The source models were trained either in a centralized or a federated setting. MME~\cite{Saito2019}, MCC~\cite{Jin2020}, SHOT~\cite{Liang2022}, and CC~\cite{Jin2024}, all of which can be computed without access to source data, were used to fine-tune the models on target data. During training, we used Cross-Entropy (CE) loss together with the losses from these domain adaptation methods, with all loss weights set to 1.

\item Transfer learning with privacy-preserved source data. ENT~\cite{Grandvalet2004}, DANN~\cite{Ganin2016}, JAN~\cite{Long2017}, CDAN~\cite{Long2018}, DAN~\cite{Long2019}, MME~\cite{Saito2019}, MDD~\cite{Zhang2019b}, MCC~\cite{Jin2020}, SHOT~\cite{Liang2022}, and CC~\cite{Jin2024} were employed, all with EEGNet as the backbone. During training, we used CE loss plus the losses from these domain adaptation methods, with all loss weights set to 1.
\end{enumerate}

\begin{figure}[htpb]\centering
	{\includegraphics[width=0.8\linewidth,clip]{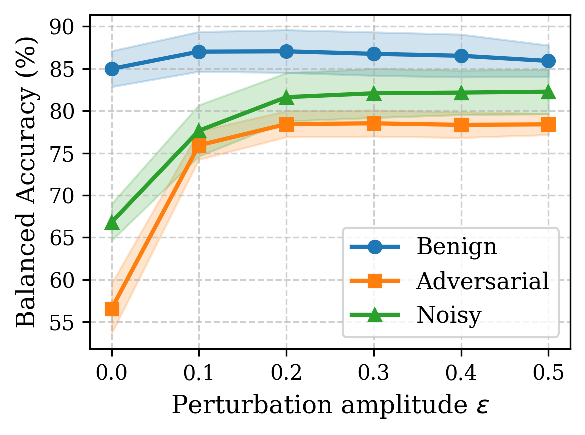}}
	\caption{$t$-SNE visualizations of BN features under centralized and federated source-free transfer scenarios on the NICU dataset.} \label{fig:pgdvar}
\end{figure}

Figure~\ref{fig:pgdvar} shows the balanced accuracy on benign, adversarial, and noisy samples for the NICU dataset with robust training perturbation amplitude $\epsilon$ varying from 0 to 0.05. Within the range of $\epsilon \in [0.01,0.05]$, the balanced accuracy on benign samples was consistently improved, while the performance on adversarial and noisy samples was substantially enhanced compared with standard training ($\epsilon=0$). This indicates that the model performance remains stable and robust across a reasonable range of $\epsilon$, consistent with observations in the existing literature~\cite{Chen2024}. Unless otherwise specified, the robust training perturbation amplitude $\epsilon$ in all robust training approaches were set to 0.03, and to 0.01 on SEED dataset~\cite{Chen2024}. Because EEG signals related to emotional responses are generally weaker and less stable than those in motor imagery and seizure detection tasks~\cite{Koelstra2011}, they are more sensitive to large adversarial perturbations.

(\ref{alg:ea1}) and (\ref{alg:ea2}) were used in all experiments to align the source data and the target calibration data for improved performance. In the test phase, we aligned the test data with the Euclidean arithmetic mean $\bar{R}$ calculated on the target domain calibration set, to avoid data leakage.

\subsection{Experiment details}

We conducted a leave-one-subject-out experiment on all datasets, where each subject served as the target user in turn, and the remaining subjects constituted the source domain.

For training the source domain model in both centralized source-free transfer and federated source-free transfer scenarios, as well as model training in the transfer with privacy-preserved source data scenario, we set the initial learning rate to 0.01, trained for 100 epochs, decayed the learning rate to 0.001 at the 50th epoch, and used a batch size of 128. During fine-tuning of the source domain model, we employed an initial learning rate of 0.001 and a batch size of 32. Given the larger volume of data in the NICU and SEED datasets, we used batch sizes of 512 (source model training) and 128 (fine-tuning stage) respectively. We referred to the original work for federated source model training~\cite{Jia2024} and generating privacy-preserved source data~\cite{Chen2024a}.

The number of steps for PGD training was 10, and the perturbation step size $\alpha=0.005$, perturbation radius $\epsilon=0.03$~\cite{Chen2024}. During federated source model training, each client, i.e., source subject, maintained independent batch normalization (BN) running mean and variance on their local data. During federated fine-tuning, the global model's BN layers' running statistics were calculated on the test batch~\cite{Jia2024}. The backbone model for all experiments was EEGNet~\cite{Lawhern2018}. More experimental details can be found in our source code \href{https://github.com/xqchen914/PAT}{https://github.com/xqchen914/PAT}.

Class-balanced sampling was applied during the training of all models to avoid bias toward the majority class.

\subsection{Main results}

The performance of various methods across different scenarios on BNCI2014001, Weibo2014, and BNCI2014002 dataset is summarized in Table~\ref{tab:main}. Under the non-transfer learning condition (denoted as ``w/o Transfer" in Table~\ref{tab:main}), where models were trained solely on source or target data, PAT achieved higher EEG classification accuracy and robustness than the baseline methods. Furthermore, across the three privacy-preserving transfer learning scenarios, PAT consistently outperformed existing methods in both EEG decoding accuracy and robustness. Additional results on the NICU and SEED datasets under the three privacy-preserving transfer learning scenarios are provided in Table~\ref{tab:appnicuseed}. PAT outperformed baseline approaches in most cases. These findings collectively demonstrate the effectiveness and adaptability of PAT for EEG decoding.

\begin{figure*}[htpb]\centering
	{\includegraphics[width=0.99\linewidth,clip]{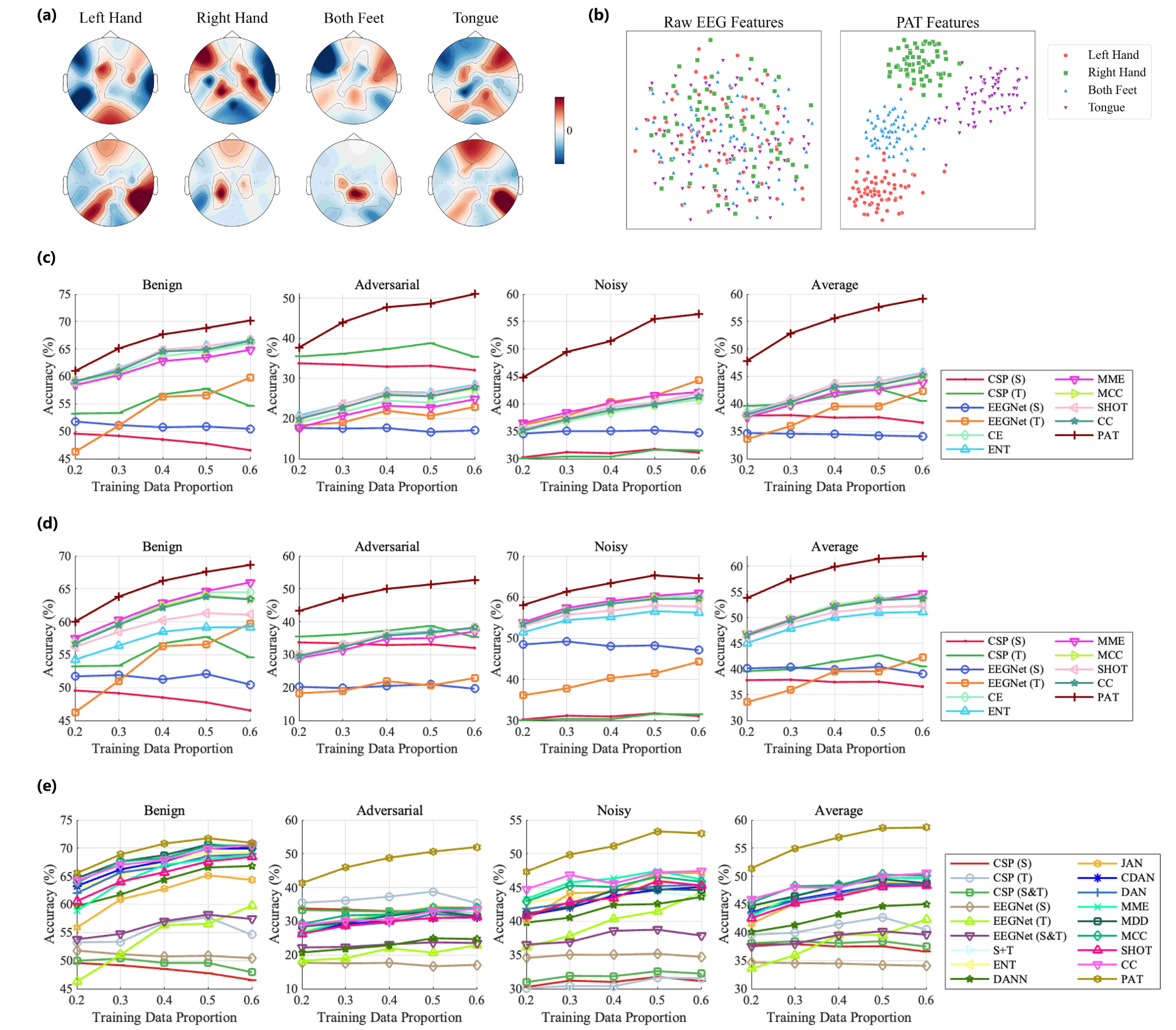}}
	\caption{Visualization of experimental results. (a) Raw EEG topography and class activation topography of PAT. (b) $t$-SNE visualizations of learned representations of PAT model. Accuracies on benign, adversarial and noisy samples with various target training data proportion on BNCI2014001 dataset (c) centralized source-free transfer learning scenario, (d) federated source-free transfer learning scenario, and (e) transfer learning scenario with privacy-preserved source data. `Average' stands for the average of `Benign', `Adversarial' and `Noisy'.} \label{fig:results}
\end{figure*}

The first row of Figure~\ref{fig:results}a presents the topography of averaged EEG trials for subject 8 from the BNCI2014001 dataset. The second row of Figure~\ref{fig:results}a depicts the corresponding channel-wise contributions in the representations learned by PAT, computed using gradient-weighted class activation mapping \cite{Selvaraju2017}. Relative to the raw EEG signals, the representations learned by PAT show more pronounced class-specific differences in the importance assigned to different channels. As shown in Figure~\ref{fig:results}b, the features learned by PAT also exhibit clearer clustering and better separability than the raw EEG features.

\begin{table*}[htbp]
  \centering \setlength{\tabcolsep}{0.4mm} \scriptsize  \centering
  \caption{Classification accuracies (\%) of various approaches. `Adv.' stands for adversarial sample. `Avg.' represents the average of `Benign', `Adv.' and `Noisy'. `Average' stands for the average results on BNCI2014001, Weibo2014 and BNCI2014002 datasets. The highest accuracies in each column of a block are marked in bold.}
    \begin{tabular}{cc|c|cccc|cccc|cccc|cccc}
    \toprule
      \multicolumn{2}{c|}{\multirow{2}[0]{*}{Setting}}   & \multirow{2}{*}{Approach}  & \multicolumn{4}{c|}{BNCI2014001} & \multicolumn{4}{c|}{Weibo2014} & \multicolumn{4}{c|}{BNCI2014002} & \multicolumn{4}{c}{Average} \\
      \multicolumn{2}{c|}{}   &   & Benign & Adv. & Noisy & Avg. & Benign & Adv. & Noisy & Avg. & Benign & Adv. & Noisy & Avg. & Benign & Adv. & Noisy & Avg. \\
    \midrule
    \multicolumn{1}{c}{\multirow{14}[4]{*}{\makecell[c]{w/o\\ Transfer}}} & \multicolumn{1}{c|}{\multirow{5}[2]{*}{\makecell[c]{Classic \\ Deep/\\Machine \\Learning}}} & CSP(T) & 55.12 & 36.61 & 30.80 & 40.84 & 40.85 & 21.25 & 24.66 & 28.92 & 73.93 & 54.71 & 59.70 & 62.78 & 56.63 & 37.52 & 38.39 & 44.18 \\
      &   & EEGNet(T) & 53.99 & 20.58 & 40.02 & 38.20 & 38.56 & 5.65 & 28.99 & 24.40 & 74.44 & 45.84 & 65.61 & 61.96 & 55.66 & 24.02 & 44.87 & 41.52 \\
      &   & CSP(S) & 48.31 & 33.08 & 31.07 & 37.49 & 35.46 & 21.11 & 31.91 & 29.49 & 70.19 & 55.75 & 58.64 & 61.52 & 51.32 & 36.65 & 40.54 & 42.83 \\
      &   & EEGNet(S) & 50.99 & 17.31 & 34.90 & 34.40 & 43.18 & 12.90 & 38.14 & 31.41 & 72.62 & 41.02 & 69.58 & 61.07 & 55.60 & 23.74 & 47.54 & 42.29 \\
      &   & EEGNet(S-Fed) & 51.49 & 20.26 & 48.24 & 40.00 & 49.64 & 16.72 & 46.52 & 37.63 & 74.63 & 47.95 & 72.57 & 65.05 & 58.58 & 28.31 & 55.78 & 47.56 \\
\cmidrule{2-19}      & \multicolumn{1}{c|}{\multirow{9}[2]{*}{\makecell[c]{Alignment- \\based \\Adversarial \\Defense}}} & ABAT & 57.49 & \textbf{43.87} & 48.31 & 49.89 & 47.65 & 33.63 & 41.93 & 41.07 & 75.40 & 63.25 & 70.51 & 69.72 & 60.18 & \textbf{46.92} & 53.58 & 53.56 \\
      &   & TRADES & 57.32 & 41.37 & 46.99 & 48.56 & 48.69 & 32.42 & 41.78 & 40.96 & 75.17 & 61.71 & 69.84 & 68.90 & 60.39 & 45.16 & 52.87 & 52.81 \\
      &   & SCR & 57.32 & 42.58 & 48.24 & 49.38 & 48.50 & 33.56 & 43.19 & 41.75 & 74.66 & 61.82 & 69.69 & 68.72 & 60.16 & 45.99 & 53.71 & 53.29 \\
      &   & AWP & 57.08 & 43.07 & 48.39 & 49.52 & 46.98 & 33.24 & 42.12 & 40.78 & 74.89 & 62.74 & 70.45 & 69.36 & 59.65 & 46.35 & 53.65 & 53.22 \\
      &   & SEAT & 44.86 & 32.67 & 34.79 & 37.44 & 45.25 & 30.63 & 38.61 & 38.16 & 66.12 & 50.81 & 57.95 & 58.29 & 52.08 & 38.04 & 43.78 & 44.63 \\
      &   & RSE & 36.50 & 25.60 & 46.23 & 36.11 & 27.09 & 11.54 & 35.11 & 24.58 & 58.28 & 50.33 & 66.95 & 58.52 & 40.62 & 29.16 & 49.43 & 39.74 \\
      &   & IT & 52.57 & 19.33 & 41.09 & 37.66 & 38.58 & 6.05 & 31.75 & 25.46 & 73.58 & 45.02 & 67.42 & 62.01 & 54.91 & 23.47 & 46.75 & 41.71 \\
      &   & SAP & 46.58 & 39.11 & 37.11 & 40.93 & 32.55 & 20.74 & 26.05 & 26.45 & 70.72 & \textbf{65.58} & 62.77 & 66.36 & 49.95 & 41.81 & 41.98 & 44.58 \\
      &   & PAT(Ours) & \textbf{63.23} & 42.48 & \textbf{52.57} & \textbf{52.76} & \textbf{54.57} & \textbf{35.04} & \textbf{47.90} & \textbf{45.84} & \textbf{78.05} & 61.02 & \textbf{71.30} & \textbf{70.12} & \textbf{65.28} & 46.18 & \textbf{57.26} & \textbf{56.24} \\
    \midrule
    \multicolumn{1}{c}{\multirow{28}[6]{*}{\makecell[c]{w/ \\Transfer}}} & \multicolumn{1}{c|}{\multirow{7}[2]{*}{\makecell[c]{Centralized\\ Source-\\Free \\Transfer}}} & CE & 62.78 & 22.97 & 38.15 & 41.30 & 59.18 & 17.60 & 47.47 & 41.42 & 76.81 & 47.86 & 72.23 & 65.64 & 66.26 & 29.48 & 52.62 & 49.45 \\
      &   & ENT & 63.45 & 25.17 & 38.88 & 42.50 & 59.39 & 18.30 & 47.88 & 41.86 & 77.67 & 51.66 & 73.34 & 67.56 & 66.84 & 31.71 & 53.37 & 50.64 \\
      &   & MME & 61.94 & 21.84 & 39.73 & 41.17 & 57.00 & 14.97 & 45.07 & 39.01 & 75.62 & 45.29 & 70.93 & 63.95 & 64.85 & 27.37 & 51.91 & 48.04 \\
      &   & MCC & 63.18 & 24.28 & 38.43 & 41.96 & 59.47 & 18.27 & 47.74 & 41.83 & 77.50 & 50.61 & 73.07 & 67.06 & 66.72 & 31.05 & 53.08 & 50.28 \\
      &   & SHOT & 63.42 & 24.92 & 38.88 & 42.41 & 59.43 & 18.19 & 47.80 & 41.81 & 77.74 & 51.49 & 73.30 & 67.51 & 66.87 & 31.53 & 53.33 & 50.58 \\
      &   & CC & 63.22 & 24.38 & 38.47 & 42.02 & 59.40 & 18.46 & 47.64 & 41.84 & 77.56 & 50.69 & 73.07 & 67.11 & 66.73 & 31.18 & 53.06 & 50.32 \\
      &   & PAT(Ours) & \textbf{66.55} & \textbf{45.81} & \textbf{51.50} & \textbf{54.62} & \textbf{64.94} & \textbf{45.51} & \textbf{57.14} & \textbf{55.86} & \textbf{77.87} & \textbf{64.89} & \textbf{75.69} & \textbf{72.82} & \textbf{69.79} & \textbf{52.07} & \textbf{61.45} & \textbf{61.10} \\
\cmidrule{2-19}      & \multicolumn{1}{c|}{\multirow{7}[2]{*}{\makecell[c]{Federated \\Source-\\Free \\Transfer}}} & CE & 61.68 & 34.50 & 57.93 & 51.37 & 59.50 & 24.16 & 55.66 & 46.44 & 79.75 & 60.57 & 77.60 & 72.64 & 66.98 & 39.74 & 63.73 & 56.82 \\
      &   & ENT & 57.49 & 34.71 & 54.77 & 48.99 & 53.15 & 25.12 & 50.44 & 42.90 & 79.63 & 61.85 & 77.54 & 73.01 & 63.42 & 40.56 & 60.92 & 54.97 \\
      &   & MME & 62.26 & 33.46 & 58.34 & 51.35 & 60.27 & 22.29 & 55.95 & 46.17 & 78.90 & 59.36 & 76.75 & 71.67 & 67.14 & 38.37 & 63.68 & 56.40 \\
      &   & MCC & 61.25 & 34.61 & 57.59 & 51.15 & 59.27 & 24.52 & 55.57 & 46.45 & 79.75 & 61.11 & 77.69 & 72.85 & 66.76 & 40.08 & 63.62 & 56.82 \\
      &   & SHOT & 59.46 & 34.77 & 56.22 & 50.15 & 57.32 & 25.58 & 54.26 & 45.72 & 79.85 & 61.71 & 77.70 & 73.09 & 65.54 & 40.69 & 62.72 & 56.32 \\
      &   & CC & 61.12 & 34.57 & 57.52 & 51.07 & 59.26 & 24.51 & 55.52 & 46.43 & 79.83 & 61.16 & 77.67 & 72.89 & 66.74 & 40.08 & 63.57 & 56.80 \\
      &   & PAT(Ours) & \textbf{65.26} & \textbf{48.93} & \textbf{62.54} & \textbf{58.91} & \textbf{64.41} & \textbf{49.06} & \textbf{61.64} & \textbf{58.37} & \textbf{80.88} & \textbf{69.35} & \textbf{79.44} & \textbf{76.56} & \textbf{70.18} & \textbf{55.78} & \textbf{67.87} & \textbf{64.61} \\
\cmidrule{2-19}      & \multicolumn{1}{c|}{\multirow{14}[2]{*}{\makecell[c]{Transfer \\with \\Privacy-\\preserved \\Source \\Data}}} & CSP(S\&T) & 49.49 & 32.94 & 31.91 & 38.11 & 37.59 & 20.59 & 32.72 & 30.30 & 69.14 & 55.28 & 60.77 & 61.73 & 52.07 & 36.27 & 41.80 & 43.38 \\
      &   & EEGNet(S\&T) & 56.22 & 23.01 & 37.74 & 38.99 & 49.85 & 18.04 & 42.48 & 36.79 & 75.68 & 48.47 & 71.45 & 65.20 & 60.58 & 29.84 & 50.56 & 46.99 \\
      &   & S+T & 68.37 & 30.69 & 43.11 & 47.39 & 58.98 & 17.65 & 50.47 & 42.37 & 79.51 & 52.48 & 74.35 & 68.78 & 68.95 & 33.60 & 55.98 & 52.84 \\
      &   & ENT & 66.83 & 29.95 & 44.14 & 46.97 & 50.08 & 11.62 & 42.74 & 34.81 & 77.28 & 52.59 & 72.07 & 67.31 & 64.73 & 31.39 & 52.98 & 49.70 \\
      &   & DANN & 63.84 & 23.00 & 41.78 & 42.87 & 55.78 & 14.76 & 49.02 & 39.85 & 77.25 & 47.82 & 72.69 & 65.92 & 65.62 & 28.53 & 54.49 & 49.55 \\
      &   & JAN & 61.81 & 31.86 & 44.57 & 46.08 & 59.47 & 19.71 & 51.96 & 43.71 & 73.22 & 54.30 & 70.41 & 65.98 & 64.83 & 35.29 & 55.65 & 51.92 \\
      &   & CDAN & 67.41 & 29.60 & 43.28 & 46.76 & 58.73 & 17.49 & 50.77 & 42.33 & 79.01 & 51.48 & 74.10 & 68.20 & 68.38 & 32.86 & 56.05 & 52.43 \\
      &   & DAN & 66.37 & 29.67 & 43.92 & 46.65 & 58.89 & 17.83 & 50.91 & 42.54 & 75.69 & 52.12 & 71.51 & 66.44 & 66.98 & 33.21 & 55.45 & 51.88 \\
      &   & MME & 65.32 & 31.25 & 45.81 & 47.46 & 51.87 & 14.18 & 44.64 & 36.90 & 79.34 & 55.61 & 74.77 & 69.91 & 65.51 & 33.68 & 55.07 & 51.42 \\
      &   & MDD & 68.37 & 30.84 & 43.47 & 47.56 & 58.83 & 17.41 & 50.22 & 42.15 & 79.18 & 52.09 & 74.42 & 68.56 & 68.79 & 33.44 & 56.04 & 52.76 \\
      &   & MCC & 68.12 & 32.15 & 45.14 & 48.47 & 54.00 & 14.45 & 46.36 & 38.27 & 79.44 & 52.85 & 73.91 & 68.73 & 67.19 & 33.15 & 55.14 & 51.83 \\
      &   & SHOT & 65.28 & 29.43 & 43.66 & 46.13 & 48.42 & 11.15 & 40.96 & 33.51 & 79.08 & 51.73 & 73.92 & 68.25 & 64.26 & 30.77 & 52.85 & 49.29 \\
      &   & CC & 67.91 & 31.19 & 46.37 & 48.49 & 53.17 & 12.89 & 45.91 & 37.32 & 79.18 & 52.00 & 74.57 & 68.58 & 66.75 & 32.03 & 55.62 & 51.46 \\
      &   & PAT(Ours) & \textbf{69.59} & \textbf{47.73} & \textbf{50.95} & \textbf{56.09} & \textbf{63.50} & \textbf{42.27} & \textbf{56.50} & \textbf{54.09} & \textbf{80.03} & \textbf{62.11} & \textbf{76.97} & \textbf{73.04} & \textbf{71.04} & \textbf{50.71} & \textbf{61.47} & \textbf{61.07} \\
    \bottomrule
    \end{tabular}%
  \label{tab:main}%
\end{table*}%

\begin{table*}[htbp]
\centering \setlength{\tabcolsep}{1mm} \scriptsize
  \caption{Balanced classification accuracies (\%) of various approaches on SEED and NICU dataset. `Avg.' column stands for the average of `Benign', `Adv.' and `Noisy'. `Average' stands for the average results on SEED and NICU. The highest accuracies in each column of a block are marked in bold.}
    \begin{tabular}{c|c|cccc|cccc|cccc}
    \toprule
  \multicolumn{1}{c|}{\multirow{2}[0]{*}{Setting}}   & \multirow{2}{*}{Approach}  & \multicolumn{4}{c|}{NICU} & \multicolumn{4}{c|}{SEED} & \multicolumn{4}{c}{Average} \\
      \multicolumn{1}{c|}{}   &    & \multicolumn{1}{c}{Benign} & \multicolumn{1}{c}{Adv.} & \multicolumn{1}{c}{Noisy} & \multicolumn{1}{c|}{Avg.} & \multicolumn{1}{c}{Benign} & \multicolumn{1}{c}{Adv.} & \multicolumn{1}{c}{Noisy} & \multicolumn{1}{c|}{Avg.} & \multicolumn{1}{c}{Benign} & \multicolumn{1}{c}{Adv.} & \multicolumn{1}{c}{Noisy} & \multicolumn{1}{c}{Avg.} \\
    \midrule
    \multicolumn{1}{c|}{\multirow{7}[2]{*}{\makecell[c]{Centralized\\ Source-\\Free \\Transfer}}} & CE & 84.96 & 56.61 & 66.87 & 69.48 & 60.98 & 26.72 & 47.04 & 44.91 & 72.97 & 41.66 & 56.96 & 57.19 \\
      & ENT & 86.06 & 58.38 & 67.66 & 70.70 & 61.10 & 27.88 & 47.15 & 45.38 & 73.58 & 43.13 & 57.41 & 58.04 \\
      & MME & 82.96 & 57.00 & 67.56 & 69.17 & 59.50 & 24.22 & 46.07 & 43.27 & 71.23 & 40.61 & 56.82 & 56.22 \\
      & MCC & 86.23 & 58.20 & 66.78 & 70.40 & 61.23 & 27.58 & 46.75 & 45.19 & 73.73 & 42.89 & 56.76 & 57.80 \\
      & SHOT & 86.21 & 58.24 & 67.49 & 70.65 & 61.24 & 27.59 & 47.12 & 45.31 & 73.72 & 42.91 & 57.30 & 57.98 \\
      & CC & 86.14 & 58.01 & 66.63 & 70.26 & 61.27 & 27.67 & 46.78 & 45.24 & 73.70 & 42.84 & 56.71 & 57.75 \\
      & PAT(Ours) & \textbf{86.76} & \textbf{78.73} & \textbf{82.14} & \textbf{82.54} & \textbf{62.04} & \textbf{38.22} & \textbf{49.35} & \textbf{49.87} & \textbf{74.40} & \textbf{58.48} & \textbf{65.74} & \textbf{66.21} \\
    \midrule
\multicolumn{1}{c|}{\multirow{7}[2]{*}{\makecell[c]{Federated \\Source-\\Free \\Transfer}}} & CE & 80.89 & 72.99 & 79.03 & 77.63 & 61.50 & 54.12 & 55.43 & 57.02 & 71.19 & 63.56 & 67.23 & 67.32 \\
      & ENT & 83.57 & 76.31 & 81.71 & 80.53 & 59.94 & 53.34 & 53.89 & 55.72 & 71.75 & 64.83 & 67.80 & 68.13 \\
      & MME & 74.83 & 63.03 & 73.60 & 70.49 & 60.73 & 52.55 & 54.29 & 55.86 & 67.78 & 57.79 & 63.95 & 63.17 \\
      & MCC & 82.98 & 75.48 & 80.97 & 79.81 & 61.51 & 54.33 & 55.66 & 57.17 & 72.25 & 64.91 & 68.31 & 68.49 \\
      & SHOT & 83.64 & 76.25 & 81.59 & 80.49 & 61.65 & 54.66 & 55.89 & 57.40 & 72.64 & 65.46 & 68.74 & 68.95 \\
      & CC & 83.00 & 75.33 & 81.12 & 79.82 & 61.63 & 54.40 & 55.81 & 57.28 & 72.32 & 64.87 & 68.47 & 68.55 \\
      & PAT(Ours) & \textbf{88.39} & \textbf{83.91} & \textbf{85.55} & \textbf{85.95} & \textbf{61.97} & \textbf{58.77} & \textbf{58.75} & \textbf{59.83} & \textbf{75.18} & \textbf{71.34} & \textbf{72.15} & \textbf{72.89} \\
    \midrule
\multicolumn{1}{c|}{\multirow{12}[2]{*}{\makecell[c]{Transfer \\with \\Privacy-\\preserved \\Source \\Data}}} & S+T & 83.94 & 49.07 & 61.07 & 64.69 & 62.22 & 31.19 & 51.56 & 48.32 & 73.08 & 40.13 & 56.32 & 56.51 \\
      & ENT & 83.97 & 56.16 & 67.10 & 69.08 & 59.07 & 33.66 & 50.69 & 47.81 & 71.52 & 44.91 & 58.89 & 58.44 \\
      & DANN & 84.74 & 52.86 & 62.82 & 66.81 & \textbf{62.61} & 28.87 & 51.45 & 47.64 & 73.68 & 40.87 & 57.13 & 57.23 \\
      & JAN & 86.77 & 70.45 & 68.41 & 75.21 & 61.81 & 34.58 & 53.62 & 50.00 & \textbf{74.29} & 52.51 & 61.02 & 62.61 \\
      & CDAN & 84.97 & 49.74 & 62.32 & 65.68 & 62.32 & 29.95 & 51.85 & 48.04 & 73.65 & 39.84 & 57.08 & 56.86 \\
      & DAN & 85.35 & 55.15 & 64.96 & 68.49 & 62.54 & 30.48 & 51.23 & 48.09 & 73.95 & 42.82 & 58.10 & 58.29 \\
      & MME & 87.05 & 74.72 & \textbf{75.91} & 79.23 & 58.08 & 39.88 & 48.80 & 48.92 & 72.56 & 57.30 & 62.36 & 64.07 \\
      & MDD & 84.50 & 49.67 & 63.34 & 65.84 & 62.51 & 30.87 & 50.97 & 48.12 & 73.50 & 40.27 & 57.15 & 56.98 \\
      & MCC & 84.52 & 59.95 & 67.29 & 70.59 & 60.93 & 33.19 & 51.09 & 48.40 & 72.73 & 46.57 & 59.19 & 59.50 \\
      & SHOT & 84.34 & 55.26 & 65.83 & 68.48 & 61.96 & 32.83 & 52.05 & 48.95 & 73.15 & 44.04 & 58.94 & 58.71 \\
      & CC & 83.97 & 54.43 & 68.31 & 68.91 & 60.58 & 31.86 & 51.03 & 47.83 & 72.28 & 43.15 & 59.67 & 58.37 \\
      & PAT(Ours) & \textbf{87.27} & \textbf{79.14} & 72.77 & \textbf{79.73} & 61.15 & \textbf{47.26} & \textbf{55.78} & \textbf{54.73} & 74.21 & \textbf{63.20} & \textbf{64.28} & \textbf{67.23} \\
    \bottomrule
    \end{tabular}%
  \label{tab:appnicuseed}%
\end{table*}%

\subsection{Performance without transfer}

The ``w/o Transfer'' row in Table~\ref{tab:main} compares conventional machine/deep learning methods and alignment-based adversarial defense methods in a non-transfer learning setting. For each method, models were trained either on source-domain data alone or on the first 20\%, 30\%, 40\%, 50\%, and 60\% of the target user's data, and the reported results are averaged over these proportions. All alignment-based adversarial defense methods were trained on target-domain data.

Although PAT yielded a slightly lower average accuracy on adversarial samples (0.74\% lower than AT), it surpassed AT by 5.10\% and 3.68\% in average accuracy on benign and noisy samples, respectively, achieving superior overall performance. These results highlight PAT's improved ability to exploit target-domain data to train EEG decoding models with higher accuracy and stronger adversarial robustness.

Moreover, alignment-based adversarial defense methods effectively enhance the accuracy and robustness of the target-domain model. For instance, AT, TRADES, SCR, and AWP achieved higher decoding accuracy and robustness than EEGNet on all three datasets. This further supports the use of alignment-based adversarial defense strategies in privacy-preserving transfer learning to improve decoding accuracy and robustness.

\subsection{Performance under centralized source-free transfer} 

Under centralized source-free transfer scenario, all source models were pre-trained in the centralized source domain data, and then fine-tuned using the first 20\%, 30\%, 40\%, 50\% and 60\% of the target user's data on three MI datasets. The average model accuracies averaged over these target-data proportions are shown in the ``Centralized Source-Free Transfer" rows of Table~\ref{tab:main}, and the detailed accuracies under different training data proportions on BNCI2014001 are shown in Figure~\ref{fig:results}c. More detailed results on Weibo2014 and BNCI2014002 are in Figures~\ref{appfig:results_weibo}a and~\ref{appfig:results_002}a. We also performed experiments on the centralized source-free transfer scenario using source data and 20\% target data for model training on NICU and SEED datasets. The results are in Table~\ref{tab:appnicuseed}.

Table~\ref{tab:main} and Table~\ref{tab:appnicuseed} show that the average accuracies of PAT on three types of samples (benign, adversarial, and noisy) were higher than the other approaches in all cases, demonstrating the effectiveness of our proposed approach.

Figure~\ref{fig:results}c shows that:
\begin{enumerate}
	\item As the amount of training data increased, the accuracies of PAT exhibited a clear upward trend across all sample types. In most cases, PAT achieved the highest classification accuracy, again demonstrating our proposed approach's effectiveness.
	\item For EEGNet(T) and various domain adaptation algorithms, increasing the amount of training data also led to noticeable accuracy improvements on benign and noisy samples. However, their accuracy on adversarial samples, i.e., adversarial robustness, may not improve much.
\end{enumerate}

\subsection{Performance under federated source-free transfer}

 Under federated source-free transfer scenario, all source models, were pre-trained under federated source domain data. Then, the first 20\%, 30\%, 40\%, 50\%, and 60\% of the target data were used for fine-tuning. The average classification accuracies averaged over these target-data proportions are shown in the ``Federated Source-Free Transfer" rows of Table~\ref{tab:main} and \ref{tab:appnicuseed}, and the detailed accuracies are shown in Figure~\ref{fig:results}d. More detailed results on Weibo2014 and BNCI2014002 are in Figures~\ref{appfig:results_weibo}b and~\ref{appfig:results_002}b. We also performed experiments on the federated source-free transfer scenario using federated source domain data and 20\% target data for model training on NICU and SEED datasets. The results are in Table~\ref{tab:appnicuseed}.

Results in the `Federated Source-Free Transfer' rows of Table~\ref{tab:main} and \ref{tab:appnicuseed} show that similar to those in the centralized source-free transfer learning scenario, the average accuracies of PAT were consistently higher than the other approaches. Figure~\ref{fig:results}d shows that PAT was always the best-performing approach.

\begin{figure}[htpb]\centering
	{\includegraphics[width=\linewidth,clip]{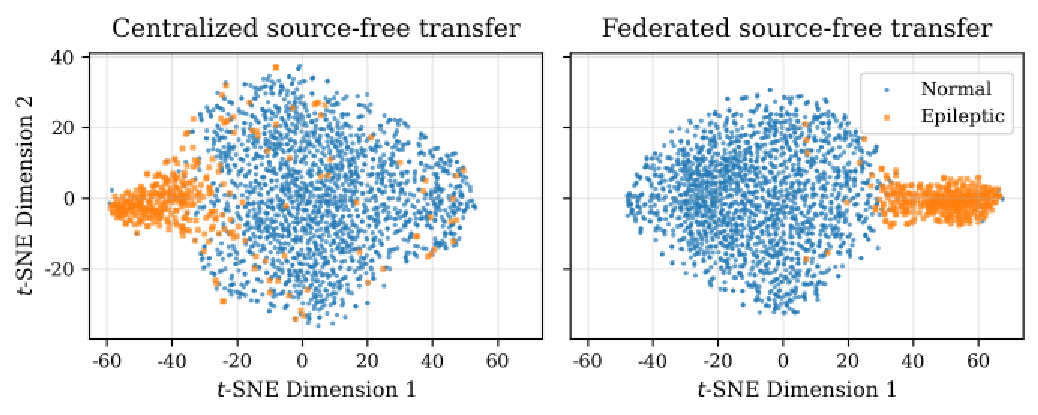}}
	\caption{$t$-SNE visualizations of BN features under centralized and federated source-free transfer scenarios on the NICU dataset.} \label{fig:bn}
\end{figure}

A comparison of the performance of EEGNet(S) and EEGNet(S-fed) in Table~\ref{tab:main} reveals that the source models trained in the federated source domain outperformed those trained in the centralized source domain. Additionally, after fine-tuning on the target domain calibration set, models pre-trained in the federated source domain achieved significantly better performance on adversarial and noisy samples than their counterparts pre-trained in the centralized source domain. This may be due to the following reasons:
\begin{enumerate}
	\item The EEG data distribution differs significantly across subjects. In federated source domain model training, each subject's BN parameters were calculated individually. This allowed the network to learn more robust global task information than computing a single set of BN parameters for all source data in centralized source domain model training.
	\item In federated source-free transfer, BN layer's running parameters were calculated for every test batch during testing, leading to better adaptation to the test data and improved robustness.
\end{enumerate}
Figure~\ref{fig:bn} shows $t$-SNE visualizations of BN features under centralized and federated source-free transfer scenarios on NICU dataset. BN features under federated source-free transfer scenarios showed more compact intra-class and separable inter-class distributions.

\subsection{Performance under transfer with privacy-protected source data}

In training, we used the perturbed privacy-preserved source data and the first 20\%, 30\%, 40\%, 50\%, and 60\% of the target data. The average classification accuracies averaged over these target-data proportions are shown in the `Transfer with Privacy-preserved Source Data' rows of Table~\ref{tab:main}, and the detailed accuracies in Figure~\ref{fig:results}e. More detailed results on Weibo2014 and BNCI2014002 are in Figures~\ref{appfig:results_weibo}c and~\ref{appfig:results_002}c. We also performed experiments on the perturbed privacy-preserved source data and 20\% target data on NICU and SEED datasets. The results are in Table~\ref{tab:appnicuseed}.

Results in the ``Transfer with Privacy-preserved Source Data" rows of Table~\ref{tab:main} and \ref{tab:appnicuseed} reveals that:
\begin{enumerate}
	\item Using source EEG data reduced performance difference between PAT and baseline approaches. Nonetheless, the proposed PAT again demonstrated better performance than others.
	\item For the CSP model, mixing source and target data (S\&T) did not always enhance the performance compared to training on source or target data alone. This may indicate that traditional models are more affected by domain distribution discrepancies, limiting their ability to capture shared features. In contrast, deep learning model EEGNet performed better with mixed source and target data.
\end{enumerate}

The classification accuracies on the benign samples are shown in the `Benign' columns of Table~\ref{tab:main} and the `Benign' subfigures of Figure~\ref{fig:results}e. Unlike models trained only on source or target domain data or a mix of the two (S\&T), domain adaptation algorithms notably improved the accuracies on benign target domain samples.

\subsection{Comparison with approaches without privacy protection}

\begin{figure*}[htpb]\centering
	{\includegraphics[width=\linewidth,clip]{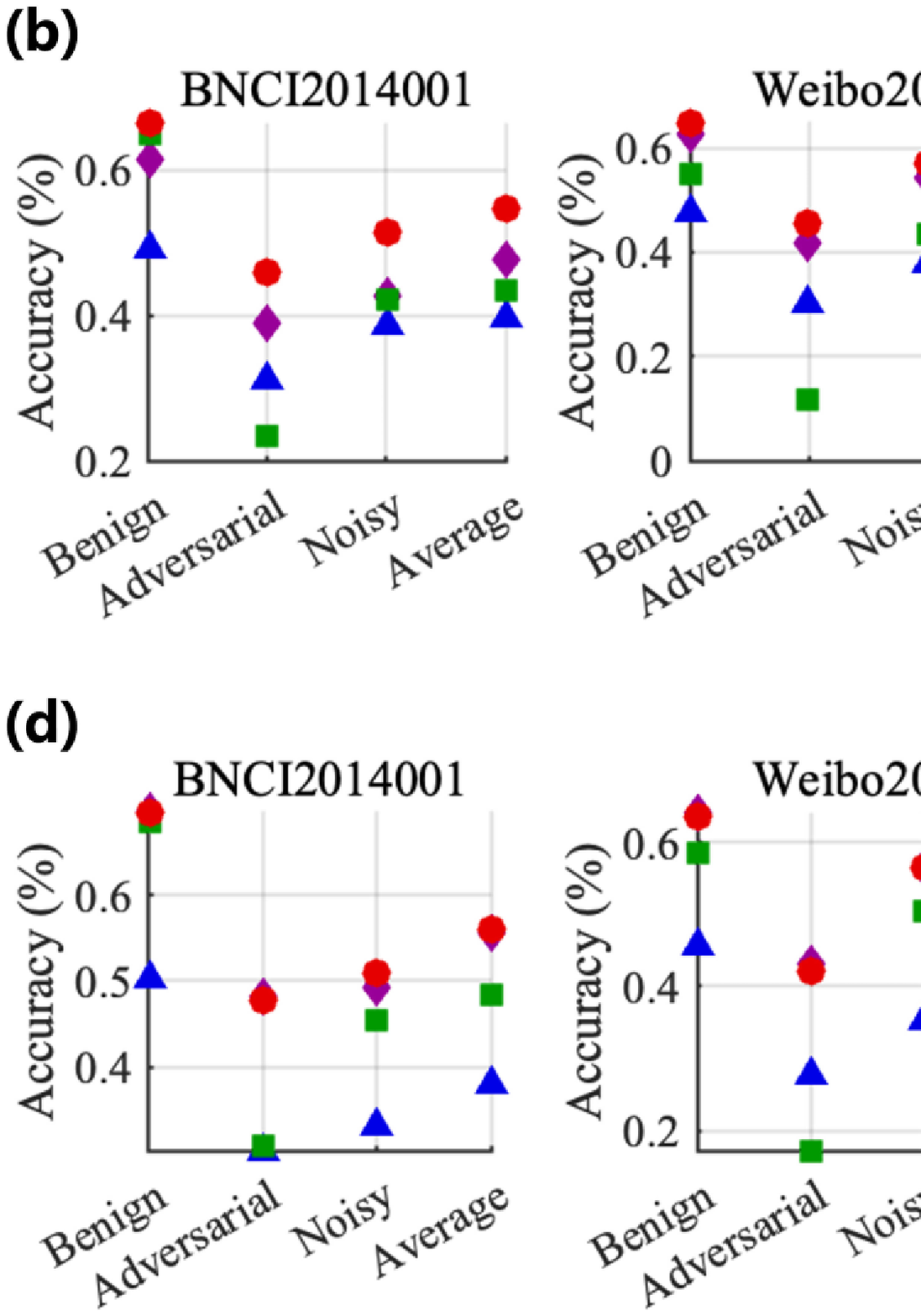}}
	\caption{More comparison results and ablation studies. (a) Comparison of PAT with transfer learning approaches without privacy protection under different training data proportions on BNCI2014001 dataset. `Average' stands for the average of `Benign', `Adversarial' and `Noisy'. Ablation studies on three datasets under (a) centralized source-free transfer scenario, (b) federated source-free transfer scenario, and (c) transfer with privacy-preserved source data.} \label{fig:tl}
\end{figure*}

To further verify the effectiveness of proposed PAT, we compared PAT in transfer scenario with privacy-preserved source data with transfer learning approaches without any privacy protection strategy. We used the source domain data and the first 20\%, 30\%, 40\%, 50\% and 60\% target domain data for model training. The average classification accuracies averaged over these target-data proportions on three MI datasets are shown in Table~\ref{tab:tl}, and the detailed accuracies on BNCI2014001 are shown in Figure~\ref{fig:tl}a. More detailed results on Weibo2014 and BNCI2014002 are in Figure~\ref{appfig:w_p}a and \ref{appfig:w_p}b. PAT again achieved the highest accuracies.

\begin{table*}[htbp]
\setlength{\tabcolsep}{1mm}
  \centering \scriptsize
  \caption{Classification accuracies (\%) of PAT under transfer scenario with privacy-preserved source data and various approaches without privacy protection. 'Adv.' stands for adversarial sample. `Avg.' represents the average of `Benign', `Adv.' and `Noisy'. `Average' column reports the average results on  `BNCI2014001', `Weibo2014' and `BNCI2014002'. The highest accuracies in each column are marked in bold.}
    \begin{tabular}{c|cccc|cccc|cccc|cccc}
    \toprule
    \multirow{2}[2]{*}{} & \multicolumn{4}{c|}{BNCI2014001} & \multicolumn{4}{c|}{Weibo2014} & \multicolumn{4}{c|}{BNCI2014002} & \multicolumn{4}{c}{Average} \\
      & Benign & Adv. & Noisy & Avg. & Benign & Adv. & Noisy & Avg. & Benign & Adv. & Noisy & Avg. & Benign & Adv. & Noisy & Avg. \\
    \midrule
    S+T & 67.60 & 23.85 & 40.73 & 44.06 & 59.41 & 14.31 & 48.54 & 40.75 & 78.74 & 44.05 & 73.32 & 65.37 & 68.58 & 27.40 & 54.20 & 50.06 \\
    ENT & 67.37 & 23.29 & 41.82 & 44.16 & 51.66 & 8.95 & 40.78 & 33.80 & 78.62 & 45.78 & 73.19 & 65.87 & 65.88 & 26.01 & 51.93 & 47.94 \\
    DANN & 62.80 & 16.56 & 39.34 & 39.57 & 55.09 & 10.57 & 45.10 & 36.92 & 76.61 & 37.77 & 71.00 & 61.79 & 64.83 & 21.64 & 51.81 & 46.09 \\
    JAN & 62.71 & 30.58 & 44.53 & 45.94 & 60.30 & 16.65 & 50.75 & 42.57 & 73.13 & 52.71 & 70.46 & 65.43 & 65.38 & 33.31 & 55.25 & 51.31 \\
    CDAN & 67.26 & 23.26 & 40.93 & 43.82 & 59.62 & 14.09 & 48.87 & 40.86 & 78.50 & 42.24 & 72.69 & 64.48 & 68.46 & 26.53 & 54.17 & 49.72 \\
    DAN & 65.86 & 23.94 & 42.42 & 44.07 & 58.90 & 14.35 & 48.76 & 40.67 & 75.79 & 48.64 & 71.23 & 65.22 & 66.85 & 28.98 & 54.14 & 49.99 \\
    MME & 65.41 & 24.86 & 42.03 & 44.10 & 52.04 & 10.14 & 41.15 & 34.44 & 79.06 & 51.92 & 74.32 & 68.43 & 65.50 & 28.97 & 52.50 & 48.99 \\
    MDD & 67.92 & 23.63 & 41.19 & 44.25 & 59.80 & 13.88 & 48.55 & 40.74 & 78.92 & 44.17 & 73.16 & 65.41 & 68.88 & 27.22 & 54.30 & 50.13 \\
    MCC & 68.65 & 25.70 & 42.58 & 45.64 & 56.12 & 12.08 & 45.45 & 37.88 & 79.19 & 45.98 & 73.35 & 66.17 & 67.99 & 27.92 & 53.80 & 49.90 \\
    SHOT & 66.97 & 23.08 & 41.49 & 43.85 & 51.97 & 9.03 & 41.40 & 34.13 & 78.81 & 44.93 & 73.34 & 65.69 & 65.91 & 25.68 & 52.08 & 47.89 \\
    CC & 68.16 & 24.39 & 42.50 & 45.02 & 54.68 & 10.39 & 42.91 & 36.00 & 79.31 & 44.80 & 73.11 & 65.74 & 67.39 & 26.53 & 52.84 & 48.92 \\
    \midrule
    PAT(Ours) &\textbf{69.59} & \textbf{47.73} & \textbf{50.95} & \textbf{56.09} & \textbf{63.50} & \textbf{42.27} & \textbf{56.50} & \textbf{54.09} & \textbf{80.03} & \textbf{62.11} & \textbf{76.97} & \textbf{73.04} & \textbf{71.04} & \textbf{50.71} & \textbf{61.47} & \textbf{61.07} \\
    \bottomrule
    \end{tabular}%
  \label{tab:tl}%
\end{table*}%

\subsection{Ablation study}

Ablation studies were conducted to validate the effectiveness of each step of PAT. Figure~\ref{fig:tl}b-d presents the results under centralized source-free transfer, federated source-free transfer, and transfer scenario with privacy-preserved source data, respectively. `DA' in Figure~\ref{fig:tl}b-d represents ``Data Augmentation", `EA' represents ``Euclidean Alignment" and `AT' represents ``Adversarial Training". `with all' stands for our complete PAT algorithm.

Data augmentation, Euclidean alignment, and adversarial training contribute substantially to improved model performance on benign, adversarial, and noisy samples. However, the contribution of data augmentation is weaker under the transfer with privacy-preserved source data setting. In Figure~\ref{fig:tl}d, the points representing `w/o DA' frequently coincide with those representing `with all', likely because the availability of source domain data depressed the improvement yielded by augmenting the target domain data.

\subsection{Computational cost trade-off}

To evaluate the computational cost of PAT, we compared its training time with six baseline approaches on the NICU dataset. Table~\ref{tab:com_time} shows the results. The baseline approaches had training times ranging from 4.78s (CE, standard EEGNet) to 11.04s (CC). PAT incurred higher training time due to the iterative PGD-based adversarial training. PAT with one PGD step was $~$3.7 times slower than CE, and PAT with 10 PGD steps was $~$14.3 times slower. This overhead is mainly caused by the generation of adversarial samples via PGD.

\begin{table*}[htbp]
  \centering \small
  \caption{Training time comparison between PAT and six baseline approaches.}
    \begin{tabular}{ccccccc|cc}
    \toprule
    Approach & CE & ENT & MME & MCC & SHOT & CC & PAT (1 step) & PAT (10 steps) \\
    \midrule
    Training Time (s) & 4.78 & 6.13 & 5.50 & 6.44 & 5.73 & 11.04 & 17.79 & 68.58 \\
    \bottomrule
    \end{tabular}%
  \label{tab:com_time}%
\end{table*}%

Despite the increased training time, PAT outperformed all baselines by 8.31\% in average accuracy and robustness, as shown in Table~\ref{tab:main}. For practical applications, the number of PGD steps can be reduced, e.g., to 1-5 steps, to lower the training time while maintaining superior performance~\cite{Chen2024}. Additionally, the inference phase of PAT has nearly identical computational cost as the standard EEGNet, as adversarial training only affects the training process and not the inference stage.

\section{Conclusions} \label{sect:CFR}

This paper has proposed PAT, a general framework for EEG-based BCIs that can be instantiated under different privacy constraints while jointly improving decoding accuracy and robustness. Experiments on five public EEG datasets and three privacy-preserving scenarios demonstrated the effectiveness of PAT. Moreover, PAT outperformed state-of-the-art transfer learning approaches that do not consider privacy protection. To the best of our knowledge, this is among the first attempts to simultaneously address three significant challenges in EEG-based BCIs: decoding accuracy, robustness, and privacy. We expect such a framework-level perspective to facilitate the design and deployment of future privacy-aware and robust EEG decoding systems in real-world applications.

\appendix
\section{Supplementary materials} \label{app1}

\subsection{Additional results} \label{app:er}

\begin{figure*}[htpb]\centering
	{\includegraphics[width=0.99\linewidth,clip]{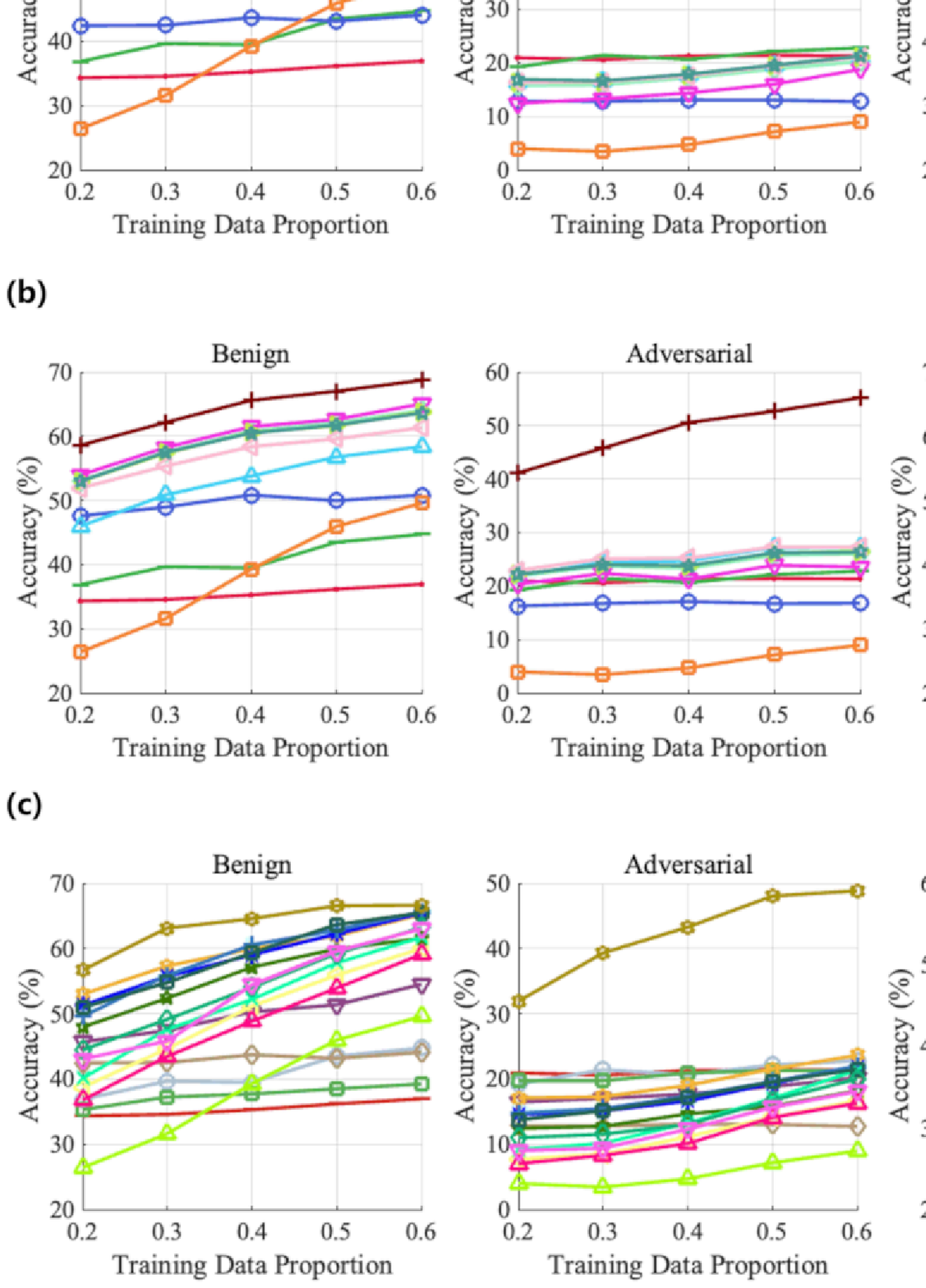}}
	\caption{Accuracies(\%) of various approaches with various target training data proportion on Weibo dataset on (c) centralized source-free transfer learning scenario, (d) federated source-free transfer learning scenario, and (e) transfer learning scenario with privacy-preserved source data. `Average' stands for the average of `Benign', `Adversarial' and `Noisy'.} \label{appfig:results_weibo}
\end{figure*}

\begin{figure*}[htpb]\centering
	{\includegraphics[width=0.99\linewidth,clip]{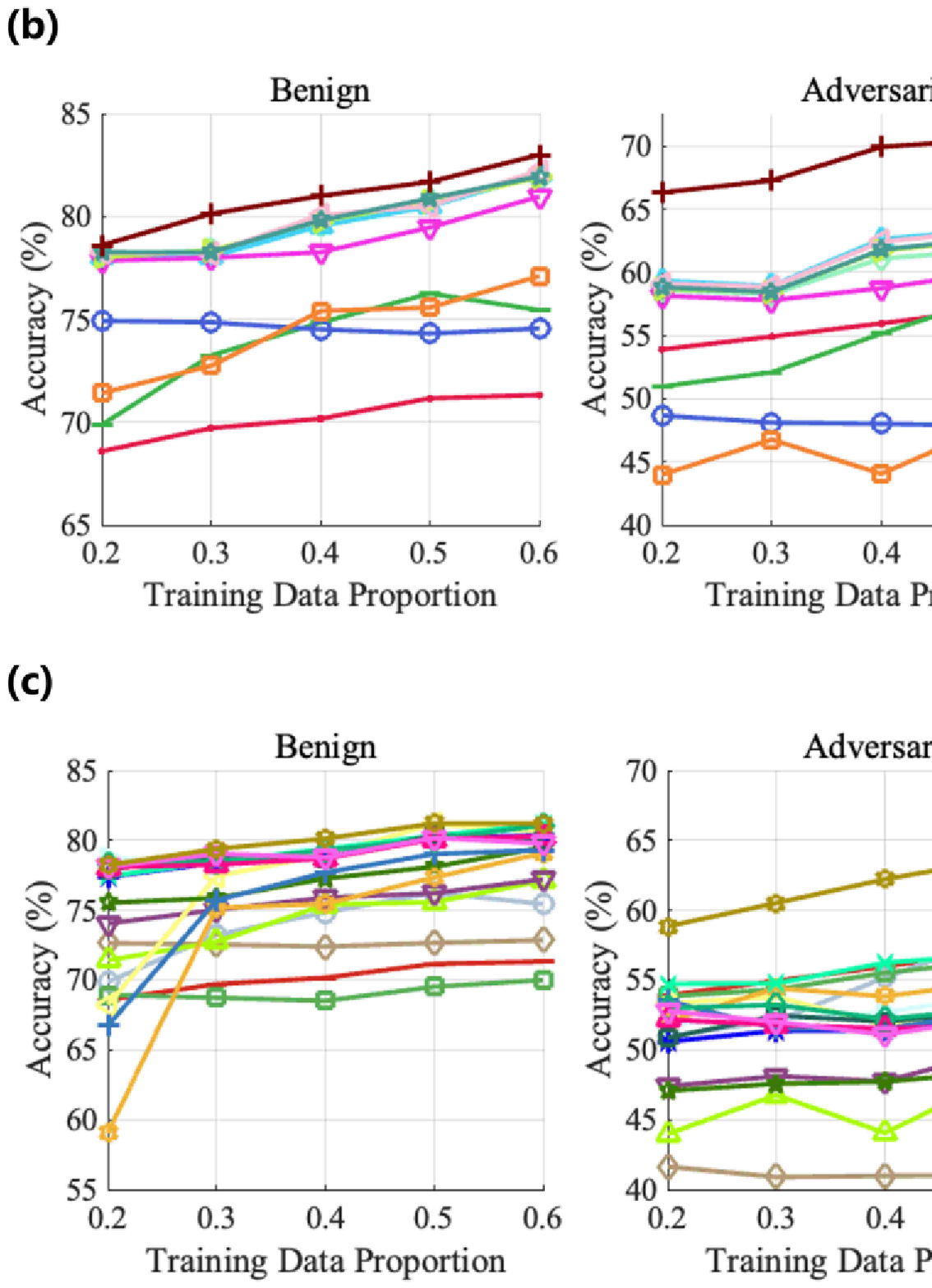}}
	\caption{Accuracies(\%) of various approaches with various target training data proportion on BNCI2014002 dataset on (c) centralized source-free transfer learning scenario, (d) federated source-free transfer learning scenario, and (e) transfer learning scenario with privacy-preserved source data. `Average' stands for the average of `Benign', `Adversarial' and `Noisy'.} \label{appfig:results_002}
\end{figure*}

Figures~\ref{appfig:results_weibo} and~\ref{appfig:results_002} present the results of various approaches under three scenarios across varying training data proportion on the Weibo and BNCI2014-002 datasets, respectively. PAT demonstrated superior accuracies in most cases.

\begin{figure*}[htpb]\centering
	{\includegraphics[width=0.99\linewidth,clip]{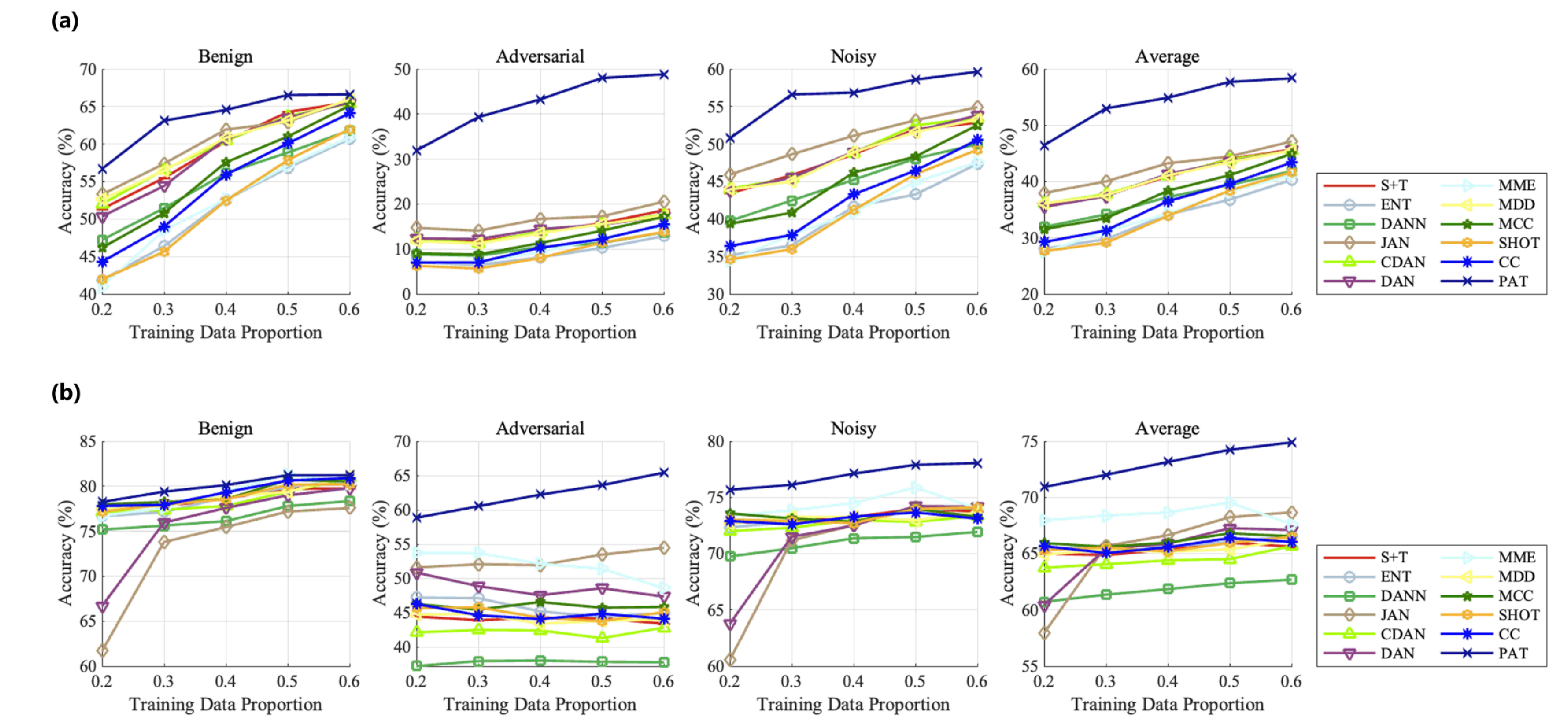}}
	\caption{Accuracies(\%) of PAT in transfer scenario with privacy-preserved source data and transfer learning approaches without privacy protection under different training data proportions on (a) Weibo and (b) BNCI2014002 dataset. `Average' stands for the average of `Benign', `Adversarial' and `Noisy'.} \label{appfig:w_p}
\end{figure*}

Figure~\ref{appfig:w_p} compares the performance of PAT in transfer scenario with privacy-preserved source data with various non-privacy-preserving transfer learning approaches on the Weibo and BNCI2014002 datasets. PAT demonstrated superior accuracy in most cases, underscoring the feasibility of accurate, robust, and privacy-preserving EEG decoding.

\end{document}